\documentclass[aps,%
twocolumn,
groupedaddress]{revtex4}

\usepackage{epsfig}

\def\e{\mbox{e}}

\def\mpc{\mbox{Mpc}}

\def\tev{\mbox{TeV}}

\def\ltap{\raisebox{-.55ex}{\rlap{$\sim$}} \raisebox{.4ex}{$<$}}
\def\gtap{\raisebox{-.55ex}{\rlap{$\sim$}} \raisebox{.4ex}{$>$}}
\def\gsim{\mathrel{\gtap}}
\def\lsim{\mathrel{\ltap}}

\begin{document}

\title{Is astronomy possible with neutral ultra-high energy
cosmic ray particles existing in the Standard Model?}
\author{P.G.~Tinyakov$^{a,c}$ and I.I.~Tkachev$^{b,c}$\\
$^a${\small\it Universit\'e Libre de Bruxelles,} \\
{\small\it CP225, Bvd du Triomphe, 1050 Bruxelles, Belgium}\\
$^b${\small\it CERN Theory Division, CH-1211 Geneva 23, Switzerland}\\
$^c${\small\it Institute for Nuclear Research, Moscow 117312, Russia
} }

\begin{abstract}
The recently observed correlation between HiRes stereo cosmic ray
events with energies $E\sim 10^{19}$~eV and BL Lacs occurs at an angle
which strongly suggests that the primary particles are neutral. We
analyze whether this correlation, if not a statistical fluctuation,
can be explained within the Standard Model, i.e., assuming only known
particles and interactions. We have not found a plausible process
which can account for these correlations. The mechanism which comes
closest --- the conversion of protons into neutrons in the IR
background of our Galaxy --- still under-produces the required
flux of neutral particles by about 2 orders of magnitude. The
situation is different at $E\sim 10^{20}$~eV where the flux of cosmic
rays at Earth may contain up to a few percent of neutrons pointing 
back to the extragalactic sources.  
\end{abstract}

\date{\today}

\pacs{PACS numbers: 98.70.Sa}

\maketitle

\section{Introduction}

In the last years it has been observed that various ultra-high energy
cosmic ray (UHECR) data sets exhibit correlations with the BL Lacertae
objects (BL Lac) at different level of significance
\cite{bllacs1,bllacs2}. Recently, the HiRes stereo data have appeared
which have unprecedented angular resolution of $\sim 0.6^\circ$. This
dataset shows correlations with BL Lacs at the angular scale
compatible with the angular resolution. The statistical significance
of the correlation is estimated to be of the order of $10^{-4}$ (11
coincidences observed at $\sim 3$ expected in the absence of
correlations) \cite{Gorbunov:2004bs,Abbasi:2005qy}.  The absence of
adjustable cuts makes it straightforward, for the first time, to
predict the signal which should be observed in the future datasets if
BL Lacs are sources of the ultra-high energy cosmic rays
\cite{predictions}.

A most striking feature of the correlation found in the HiRes data
is that it occurs at an angle which is much smaller than the typical
deflection of a proton of corresponding energy in the Galactic
magnetic field (GMF). The purpose of this paper is to investigate whether
the existence of such correlations can be explained within the
Standard Model, i.e., assuming only known particles and
interactions. We will argue that this is extremely unlikely, if not
impossible.

In order to proceed with the argument we need to make several
assumptions. Although these assumptions are plausible, they may not
be valid. If this be the case, the results of our analysis should be
reconsidered.

The assumptions are as follows:
\begin{itemize}
\item[i)] The fraction of correlating events at energy $E>10^{19}$~eV
  is larger than $\sim 1$\%.

\item[ii)] The Galactic magnetic field around Earth location has a
  coherent component with the strength of order $2-3~\mu G$.

\item[iii)] The distances to BL Lacs which are counterparts (sources) of
  correlating events are larger than $\sim 100$~Mpc.
\end{itemize}

The validity of the assumption i) has been discussed in detail in
Refs.~\cite{predictions}. Note that it is implicitly assumed
here that energies of cosmic rays are measured correctly.

The assumption ii) is the widely accepted value of the Galactic
magnetic field in the vicinity of the Earth (for recent reviews see,
e.g., Refs.~\cite{GMF}). The precise magnitude of MGF is not important
for the argument; its variations by a factor 2-3 would not change our
conclusions.

Finally, the assumption iii) is needed because some of the BL Lacs
which contribute to correlations have unknown redshifts. It
is usually expected that these redshifts exceed $0.1-0.2$. 

Given the assumptions i)--iii), the argument proceeds as follows.  The
deflection of a $E=10^{20}$~eV proton in the $2~\mu$G coherent field
extending over 1 kpc is $1^\circ$. Most of the events, however, have
much lower energies (for the events of energies $E>10^{19}$~eV with
the spectrum falling like $\sim 1/E^3$ the median energy is $1.5\times
10^{19}$~eV). Since the correlating events follow the same
distribution \cite{Abbasi:2005qy}, their typical deflections would be
$\gsim 7^\circ$. The correlation with the sources would therefore be
destroyed. At such a small angular scale as observed, the correlations
can survive in the following cases only:
\begin{itemize}
\item[1)] There exist ``windows'' in the Galactic magnetic field with
a very low value of the coherent component.
\item[2)] A fraction of primary particles is neutral.
\item[3)] A fraction of primaries is converted to neutral particles
before entering the Galactic magnetic field, i.e., at least 1~kpc from
the Earth (assuming the Galactic magnetic field does not extend
further than $\sim 1$~kpc from the disk).
\end{itemize}
We consider these three possibilities in Sects.II-V. In this paper we
limit ourselves to mechanisms which are based on particles and
interactions existing in the Standard Model. We show that none of such
mechanisms can explain the observed correlation, unless very unlikely
assumptions are made. In the last Sect.VI we summarize the arguments
and present the conclusions.

\section{Magnetic fields}
\label{sec:magnetic-fields}

\subsection{Galactic magnetic field}

The Galactic magnetic field (GMF) consists of two components, the
coherent and the turbulent one. The existence of the coherent
component is the main reason why the UHECR-BL Lacs correlations at $E
\sim 10^{19}$ eV cannot be explained by protons. In models which are
currently in use, the coherent GMF extends to the whole Galaxy, being
described by a simple analytic function. However, such a picture is
probably an oversimplification. Observationally, there are many
anomalies and features in the Galactic magnetic field. It is not
totally excluded that the coherent component is ``patchy''. In other
words, there may exist ``windows'' where the coherent component is
negligible. In this case the ultra-high energy protons may cross the
GMF undeflected when they come from the directions of these
``windows''. One may thus try to explain the observed correlations by
the existence of such windows.

For this mechanism to work the random component of the GMF in windows
also has to satisfy some requirements. The deflections of protons in
the random field is estimated as follows
\begin{equation}
\delta_r = 0.5^\circ\cdot \left({10^{19}~{\rm eV} \over E} \right)
\left({B_r \over 4~\mu{\rm G}}\right) \sqrt{D\over 1~{\rm kpc}}\,
\sqrt{L_c\over 1~{\rm pc}}\; , 
\label{FstEstimeate}
\end{equation}
where $E$ is the energy of proton, $B_r$ and $L_c$ are the rms value
and the coherence length of the random magnetic field, respectively,
while $D$ is the propagation distance. This deflection has to be
(much) smaller than $0.5^\circ$. 

The coherence length $L_c$ is the most uncertain of the above
parameters. Quite often a large values of $L_c$ up to $L_c \sim 50$~pc
are assumed. On the contrary, in those regions of the sky where the
spectrum of the magnetic field fluctuations was measured, $L_c$ turns
out to be small \cite{Tinyakov:2004pw}.  For instance, in
Ref.~\cite{Gaensler:2000gz} the linearly polarized continuum emission
was studied in the test region near the Galactic plane covering the
range $325.5^\circ < l < 332.5^\circ$, $-0.5^\circ < b <
3.5^\circ$. Polarized emission was found to originate mainly at the
distance of $\sim 3.5$ kpc. Interestingly, two large areas of a few
square degrees each were found to be devoid of polarization. It was
argued that these voids were produced by the foreground in which the
magnetic field is disordered, with the coherence length being $L_c
\sim 0.1 - 0.2$ pc. In these voids, the projection of the coherent
component of the magnetic field on the line of sight was found to be
$< 0.15$ of the rms value of the random field strength. In the rest of
the test region, i.e. outside of the voids, the coherence length is
much larger, but still the outer scale of turbulence did not exceed
$2$~pc \cite{Haverkorn:2004fw}. Thus, the existence of regions with
$\delta_r<0.5^\circ$ does not seem impossible.

This mechanism has a specific signature which is straightforward to
test. If there exist ``windows'' with the small coherent component
of GMF, the Faraday rotation measures must be small in these windows
as well. In other words, the Faraday rotations in the directions of
correlating UHECR events must be anomalously small. This may be tested
statistically by comparing the distribution of Faraday rotations in
the direction of {\em correlating} events with the distribution of
Faraday rotations in the random directions selected according to the
distribution of BL Lacs and {\em all} cosmic ray events\footnote{One
may construct this set by choosing the directions to BL Lacs
correlating with the {\em Monte Carlo simulated} cosmic ray events. In
this way the distributions of both BL Lacs and the cosmic ray events
are taken into account.}. We have performed this test with the
existing data and found that the two distributions are indeed
different (Faraday rotations in the directions selected with real data
are anomalously small) with the significance of $\sim 4\%$ according
to the Kolmogorov-Smirnov test. This is not a very significant
deviation. The result demonstrates, however, that the method may work
quite well with the future larger datasets.

Although the existence of ``windows'' in the coherent component of the
Galactic magnetic field goes against the standard lore, a much better
understanding of the Galactic magnetic field is required to definitely
rule it out. 

\subsection{Extragalactic magnetic fields}

For the mechanism outlined above to work, the 
extragalactic magnetic fields
have to satisfy certain requirements (which also apply to the
scenarios
considered in Sec.~\ref{sec:conversion}). 
The extragalactic magnetic fields are not measured. Computer
simulations indicate \cite{Dolag:2003ra,Bruggen:2005ti} that the magnetic field
strength in voids between clusters can be very small, $B_r < 10^{-12}~{\rm
G}$, while the coherence length can easily be significantly smaller
than 1 Mpc. Eq.~(\ref{FstEstimeate}) then shows that the deflections in voids
are negligible. It is interesting to note that EGMF with such small magnitude
are in principle measurable  in observations of TeV gamma-rays from 
distant blazars \cite{Neronov:2006hc}.

The strength of the field in filaments is
larger. However, the probability to cross many filaments is small
and regions with small deflections can occupy rather large
fraction of the sky area
\cite{Dolag:2003ra,Bruggen:2005ti} (see however \cite{Sigl:2003ay}).  
Overall, the model where the extragalactic
magnetic fields are sufficiently small and do not spoil correlations
is acceptable at present.

\section{Neutral primaries}

Among known neutral particles the following are sufficiently stable to
propagate over extragalactic distances: neutrino, photon and atoms. In this
section we discuss the possibility to explain correlations by assuming
that primary cosmic rays are composed of these particles.

Both neutrino and photon initiate air showers deeper in the atmosphere
than the hadronic primary particles.  Therefore, these models can be
falsified with already existing data by, e.g., comparing the $X_{\rm
max}$ distributions of the correlating events with that of the whole
set. Since the corresponding data are still unpublished, we briefly
discuss the models based on neutrino and photon and show that they
have difficulties {\it per se}, even without referring to $X_{\rm
max}$ distributions.


\paragraph{\bf Neutrino.}

At $E \gsim 10^{19}$ eV the neutrino cross-section with protons is
smaller by a factor of $\sim 3\times10^{-7}$ than the $pp$
cross-section \cite{Gandhi:1998ri}.  Therefore, the optical depth of
the atmosphere for neutrinos is $3\times10^{-5}$.  On the other hand,
at this energy the neutrino flux cannot exceed the flux of hadronic
cosmic rays by more than a factor of $50$ \cite{Kalashev:2002kx}.  It
follows that at most $\mbox{(a few)}\times 10^{-4}$ of all cosmic ray
events can be due to neutrinos. This is more than a factor of 10 lower
than needed to explain correlations. Thus, neutrino with the
standard weak interactions cannot explain correlations observed in
the HiRes data set.

A ``genuine'' (hypothetical) neutrino mechanism would involve strong
neutrino interactions with the atmosphere at high energies
\cite{strong-neutrino-int}. As such a behavior in not a part of the
Standard Model, the corresponding speculations fall outside of the
scope of the present paper.

Another possibility which exists within the minimal extention of the
Standard Model by the non-zero neutrino masses, the $Z$-burst
mechanism \cite{Zburst}, requires an unnaturally large flux of
neutrinos at $E > 10^{22}$ eV which is in conflict with the limits on
neutrino flux from radio experiments \cite{Semikoz:2003wv}. The
particles observed at Earth in this mechanism are mostly photons
produced in the interactions of UHE neutrinos with the cosmological
neutrino background on their way to the Earth. Low radio-background
and small values of EGMF are required to avoid the conflict with the
upper bound on the diffuse flux of gamma rays \cite{Kalashev:2001sh}.

\paragraph{\bf Photon.}

A set of conditions under which the UHE photons can reach the Earth from BL
Lacs was considered in Ref.~\cite{Kalashev:2001qp}. On their way the photons
interact with the CMBR and radio background photons producing $e^+e^-$
pairs, one of these particles typically carrying most of the energy.
These leading particles in turn Compton up-scatter
CMBR photons to the energy almost equal to the energy of the original
photon. This process is usually
referred to as the electromagnetic cascade. 
The developing electromagnetic cascade
can reach the Earth from several hundred megaparsecs 
with energy $E \sim 10^{19}$ eV if the following
conditions are satisfied:
\begin{itemize}
\item the radio-background is small, smaller than the theoretically
expected value;
\item the injection spectrum $\propto E^{-\alpha}$ is hard, $\alpha
      \lsim 1.5$;
\item maximum energy of photons at the source reaches $10^{23}$ eV
\item EGMF are small, $B < 10^{-12}$ G;
\item sources are predominantly photonic, $N_\gamma/N_p \gsim 10^2$.
\end{itemize}
These conditions impose extreme requirements on the astrophysical
sites where such photons can be produced. There are no candidates
known which could satisfy these requirements. 

\paragraph{\bf Atoms.}

It may, in principle, happen that in the cosmological radiation field
the proton produces an $e^+e^-$-pair and ``dresses'' itself with the
electron forming an hydrogen atom and emitting a free positron. The
differential cross section of electromagnetic pair production by a
single photon in the Coulomb field of a nucleus with subsequent
capture of the electron is estimated as \cite{Aste:1995sc}
\[
\frac{d\sigma}{dE_p} = {4\pi \alpha^6 Z^5 \over m_e^2} {1\over E_p},
\]
where $Z$ is the electric charge of an ion and the positron energy
$E_p$ is supposed to be much larger than $m_e$. Multiplying the cross
section integrated over energy by the density of the CMB photons one
may estimate the rate of the formation of hydrogen atoms, $Z=1$, as
\[
R_{\rm formation} \sim 10^{-5}\; \mpc^{-1}.
\]
The decay rate (ionization on the CMB radiation) is estimated in a
standard way by using the Klein-Nishina cross section. One finds
\[
R_{\rm decay} \sim 100\; \mpc^{-1}.
\]
Thus, the fraction of neutral particles (atoms) produced by this
mechanism is of the order of $10^{-7}$, which is too small to explain
correlations.

As a side remark note that for a heavy nuclei the rates of radiative capture
and ionization are comparable when $Z \approx 25$. This corresponds
to the typical equilibrium charge of a heavy ion (iron or heavier) 
propagating in the CMBR.

\section{Conversion to neutrons in or near the Galaxy}
\label{sec:conversion}

In order to be able to fly over 1~kpc (the thickness of the Galactic
magnetic field) a neutral particle created at the outskirts of the
Galaxy has to be sufficiently stable. At energy $10^{19}$~eV this
implies for the rest-frame lifetime
\[
\tau_0 > 10\, {\rm s}~ \left({m\over 1\;{\rm GeV}} \right) \left( {
10^{19}\;{\rm eV} \over E }\right),
\]
where $E$ and $m$ are the energy and the mass of the particle,
respectively. Among known particles which we have not yet discussed
only neutrons satisfy this requirement. In this section we consider
various mechanisms of neutron creation in or near the Galaxy.

There are several ways to produce neutrons in the Standard Model:
photodisintegration of nuclei, photoproduction on background photons
by protons, creation in $pp$ reactions and in inverse $\beta$-decay on
background neutrinos or photons. We consider these mechanisms in turn
and argue that none of them can produce a sufficient fraction of
neutrons in the cosmic ray flux.

\subsection{Inverse $\beta$-decay on background neutrinos.} 

The simplest of the above mechanisms is the inverse beta-decay, $p +
\bar\nu\to n+e^+$. The cross section of this reaction is \cite{okun}
\[
\sigma(p\bar\nu \to n e^+) \simeq {1\over \pi} G^2_F (g_V^2 +
3g_A^2)E^2,
\]
where $g_V^2 + 3g_A^2\sim 5.7$ and $E$ is the energy of neutrino in
the proton rest frame. When $E$ reaches $\sim 1$~GeV the cross section
levels out and stabilizes at the value of $\sigma_{\rm max} \sim
10^{-14}$~barn. With this maximum value taken for the estimate, the
rate of the conversion is
\begin{equation}
R_{\rm max} \sim 4\times 10^{-12}~ {\mpc}^{-1}. \label{eq:weak-rate}
\end{equation}
Thus, these processes are totally negligible.

\subsection{Creation of neutrons in the radiation fields.}

The process of creation of neutrons in interactions of CR primaries
with the background photons produces the largest contribution,
therefore we consider it in most detail.

\subsubsection{Galactic and extragalactic 
radiation fields and reaction rates.}

In the laboratory frame, the rate of reactions with the photon
background is given by the standard expression,
\begin{equation}
R = \int d^3p\, n({\bf p}) (1-v\cos\theta) 
\sigma(\tilde\omega),
\label{eq:rate-generic}
\end{equation}
where $n({\bf p})$ is the photon density in the laboratory frame,
$\sigma(\tilde\omega)$ is the cross section of the relevant reaction
in the rest frame of the primary particle as the function of the
energy of the incident photon $\tilde\omega=\gamma p(1 -
v\cos\theta)$, $\gamma$ is the gamma-factor of the primary particle in
the laboratory frame and $v$ is its velocity ($\gamma =
1/\sqrt{1-v^2}$). In what follows we assume $\gamma\gg 1$.

In the case of the isotropic background this expression may be
simplified. Integrating over angles we find
\begin{equation}
R(\gamma) = {2\pi\over \gamma^2} \int_0^\infty dp \, n({\bf p})
\int_0^{2\gamma p} d\omega\,\omega \sigma(\omega).
\label{eq:rate_iso}
\end{equation}
For the  black body radiation with
temperature $T$ one has
\begin{equation}
n({\bf p}) = n_{\rm T}(p) \equiv {2\over (2\pi)^3} {1\over \exp(p/T) -1}.
\label{eq:n_bb}
\end{equation}
This gives the answer in the case of CMBR. Other backgrounds, Galactic and
extragalactic, are usually characterized in the literature by the spectral
energy distribution $I(\nu,{\bf i})$ (energy per unit frequency per unit solid
angle) which in turn is usually quoted in terms of the Planck function $B_\nu
(T)$ and emissivity $\epsilon$
\begin{equation}
I(\nu,{\bf i}) = \epsilon(\nu,{\bf i})\, B_\nu (T).
\label{eq:Inu}
\end{equation}
Here ${\bf i}$ stands for the unit vector in the direction of observation.
For the black body radiation $\epsilon(\nu,{\bf i})=1$.
The Planck function, being written as a function of photon momentum $p =
2\pi\nu$, takes the form 
\begin{equation}
B_p (T) = p^3 n_{\rm T}( p).
\label{eq:Bnu}
\end{equation}
Therefore, the photon number density for the background 
with the known emissivity is
given by the expression
\begin{equation}
n({\bf p}) = \epsilon({\bf p}/2\pi)\; n_{\rm T}^{~}(p).
\label{eq:n_generic}
\end{equation}

In what follows we will be interested in the Galactic and
extragalactic far-infrared backgrounds (FIRB) 
(for a recent review see \cite{IR}). According to
Ref~\cite{Fixsen:1998kq}, the isotropic extragalactic FIRB can be
parameterized by
\begin{equation}
\epsilon(p) = 1.3 \times 10^{-5}\, (p/p_0)^{0.64} ,
\label{eq:eps_extragal}
\end{equation}
where $p_0 = 144$ K (which corresponds to $\nu_0= 100\; {\rm
cm}^{-1}$), while the temperature parameter in $ n_{\rm T}( p)$
corresponds to $T=18.5$ K.
 
The Galactic FIRB has been measured by COBE/DIRBE. The spectral energy
density $I(\nu,{\bf i})$ as a function of galactic coordinates can be
downloaded from \cite{COBE}. The
radiation is dominated by the Galactic plane where the Galactic bulge
is by far the brightest region. One may approximate this radiation
field by a point source in the Galactic center. We have verified that
this approximation gives a good agreement with the exact calculations
for cosmic ray trajectories which do not pass close to the Galactic
center.

According to \cite{wright-et-al}, the averaged spectral properties of
the Galactic FIRB can be described by $n_{\rm T}( p)$ with $T=20.4$
K and $\epsilon(p) \propto p^2$. Therefore, in what follows for the
Galactic FIRB we use
\begin{equation}
\epsilon({\bf p}) = {I_0\over r^2}\; p^{2}\; \delta({\bf n-n}_0),
\label{eq:eps_gal}
\end{equation}
where $I_0$ is the normalization factor, ${\bf n}_0$ is the unit
vector in the direction from the Galactic center, and $r$ is the
distance to the Galactic center. The constant $I_0$ can be found by
normalizing the total luminosity within the Sun orbit to the measured
value $L_G = 1.8\times 10^{10}~L_\odot\sim 7\times 10^{36}$~W
\cite{wright-et-al}. We find
\[
I_0 = {63 L_G\over 8 \pi^4 T^{6}}.
\]

The reaction rate Eq.~(\ref{eq:rate-generic}) in this case can
be expressed as
\begin{equation}
R(\gamma,r,\theta) =
\frac{126}{64\pi^7}\frac{L_G}{T^6r^2}(1-\mu)\int_0^\infty \frac{dp\,
p^4\, \sigma(\tilde\omega)}{\exp (p/T)-1}.
\label{eq:rate-point}
\end{equation}
Here $\tilde\omega \equiv \gamma p(1 - \mu)$, $\mu \equiv\cos\theta$
and $\theta$ is the collision angle between the CR primary and the
background photon.

\subsubsection{Conversion in the extragalactic space.}

The fraction of neutrons created over the distance $dl$ is $Rdl$.  Due
to the finite neutron lifetime the fraction of neutrons which
reach the solar system is given by
\begin{equation}
F(\gamma) = R \int_0^\infty \e^{-l/\lambda} dl = R\lambda,
\label{eq:cosmoP-rate}
\end{equation}
where $R$ is given by the expression Eq.~(\ref{eq:rate_iso})
and $\lambda$ is the mean propagation distance of the
free neutron,
\[
\lambda = 0.86\, \frac{\gamma}{10^{11}}\, {\rm Mpc}.
\]

\begin{figure}
\hspace{-.5cm}\includegraphics[width=.5\textwidth]{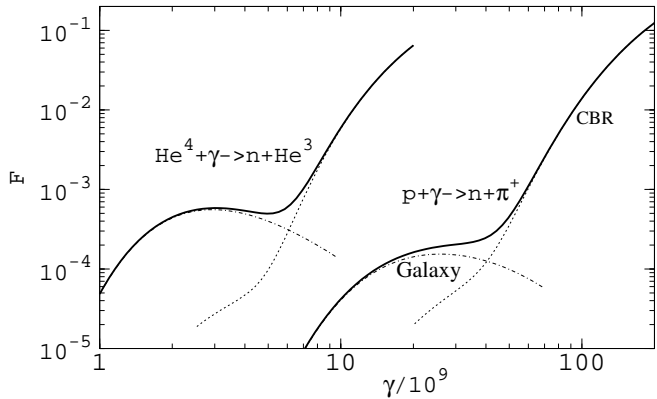}
\caption{The fraction $F$ of neutrons produced per one incident
  particle (solid lines) in the reactions $^4{\rm He} +\gamma
  \rightarrow\, ^3{\rm He} + n$ (left curve) and $p+\gamma \rightarrow
  n + \pi^+$ (right curve) on the background radiation fields as a
  function of the $\gamma$-factor of the incident particle. Dotted and
  dashed-dotted lines show contributions of the extragalactic and
  Galactic backgrounds, respectively.}
\label{fig:F}
\end{figure}

The function $F(\gamma)$ is shown in Fig.~\ref{fig:F} by dotted lines
for the two reactions, the pion photoproduction $p+\gamma \rightarrow
n + \pi^+$ (right curve) and the reaction of nuclear photodissociation
$^4{\rm He} +\gamma \rightarrow\, ^3{\rm He} + n$ (left curve). In
these calculations the experimentally measured cross sections of the
corresponding reactions were used \cite{Fujii:1976jg,photodis}.

\subsubsection{Conversion in the Galactic infrared radiation field.}

In this case the number of neutrons produced per one incident particle
is determined by the rate (\ref{eq:rate-point}) integrated along the
particle trajectory,
\begin{equation}
F(\gamma,\psi) = \int_0^\infty dl\, R(\gamma,r,\theta)\,  \e^{-l/\lambda},
\label{eq:integral-along-traj}
\end{equation}
where $l$ is the distance from the Sun along the trajectory and $r$ is
the distance from the current point to the Galactic center. In the
case when the radiation field is approximated by the single source in
the Galactic center, the particle trajectory is completely
characterized by the angle $\psi$ which it forms with the direction to
the Galactic anti-center ($\psi=\pi$ corresponds to the trajectory
which passed through the Galactic center). In terms of this angle the
distance $r$ entering Eq.~(\ref{eq:integral-along-traj}) reads
\[
r = \sqrt{D^2 + l^2 +2 D l\cos\psi}, 
\]
while the collision angle $\theta$ is 
\[
\cos\theta = - {D\cos\psi +l\over r},
\]
where $D\approx 8$ kpc is the distance from the Sun to the Galactic
center

The Galactic contribution $F(\gamma,\psi)$ to the fraction of the
produced neutrons in the case $\psi =90^\circ$ 
is shown in Fig.~\ref{fig:F} by the dashed-dotted lines for the
reactions $p+\gamma \rightarrow n + \pi^+$ (right curve) and $^4{\rm
He} +\gamma \, \rightarrow\, ^3{\rm He} + n$ (left curve).  Here again
we have used the cross sections measured experimentally.

As far as the correlations observed in the HiRes data at $E>10^{19}$ eV 
are concerned, the
relevant range of the $\gamma$-factors is $(1-2)\times 10^{10}$. In
this region the reaction $^4{\rm He} +\gamma \, \rightarrow\, ^3{\rm
He} + n$ is irrelevant for distant sources. Indeed, in the case 
of $^4{\rm He}$ these
$\gamma$-factors correspond to energies $(4-8)\times 10^{19}$~eV. The
helium nuclei of such energies do not propagate over several hundred
megaparsecs \cite{nuclei-prop}, so they cannot be present in the cosmic ray
flux coming from BL Lacs. The other reaction, $p+\gamma \rightarrow n
+ \pi^+$, produces a fraction of neutrons at the level of $\mbox{(a
few)}\times 10^{-4}$ (see Fig.~\ref{fig:F}), which is not sufficient
to explain correlations by almost two orders of magnitude.

\subsection{Neutron production in collisions with interstellar matter.}

Neutrons can be produced in collisions of hadronic primaries with the
interstellar gas in the galaxy. The conversion probability is given by
the optical depth $\tau = {\cal N} \sigma_{g}^{~}$, where ${\cal N}$
is the column density of the intervening interstellar gas in a given
direction and $\sigma_{g}$ is the interaction cross section. In order
to explain correlations \cite{Gorbunov:2004bs,Abbasi:2005qy} one needs
$\tau \gsim 10^{-2}$.

A typical value of the HI (neutral hydrogen) column density in
directions of the Galactic poles is ${\cal N}_{\rm HI} \approx
10^{20}~\mbox{cm}^{-2}$ \cite{Dickey:1990mf}.  Using as an upper limit
for $ \sigma_{g}$ the value of the total $pp$ cross-section at
relevant energies, $\sigma_{pp} \approx ~100~ {\rm mb} =
10^{-25}~\mbox{cm}^{2}$, one finds  $\tau_{pp} \sim 10^{-5}$, which
is too small to produce the required fraction of neutrons.

The argument can be rephrased in a different way. One may assume that
a mass fraction $\eta$ of the Galactic halo consists of baryons
including nuclei, neutral and ionized gas and possibly dark baryons.
The column mass density of matter in the direction of the Galactic
anti-center, as deduced from the Milky Way rotational curve, is $\sim
10^{22}\;{\rm GeV\;cm}^{-2}$ \cite{Boyarsky:2006fg}, and therefore the
column density of baryons is $\sim \eta\, 10^{22}~\mbox{cm}^{-2}$.  To
reproduce the required rate of $pn$ conversions one would need a
fraction $\eta \gsim 10$, which is clearly impossible.

As a side remark let us point out that neutrons can be, in principle,
produced in the interactions of primary protons with a non-baryonic
dark matter in the Galactic halo. Parameterizing the relevant
cross-section in the energy range of interest as $\sigma \equiv
E_0^{-2}$ and making use of the matter column density of the Galactic
halo cited above we find
\[
\tau_{p\rm DM}^{~} \sim 10^{-2}\; \left(\frac{1\;\tev}{E_0}\right)^2\;
\left(\frac{1\;\rm eV}{m_{\rm DM}^{~}}\right),
\]
where $m_{\rm DM}$ is the mass of the dark matter particle. Among the
scenarios involving new physics, this one has several advantages. It
automatically provides a normal shower development in the atmosphere
(contrary to the models  with new particles as 
neutral messengers \cite{Chung:1997rz,Berezinsky:2001fy}) and
avoids the problem of messenger production in AGNs \cite{Kachelriess:2003yy}. 
In addition, we
know from precision cosmological data that the non-baryonic dark
matter must exist. Correlations in this scenario should dissappear at
$E \lsim 10^{17}$~eV due to the final life-time of the neutron.  Note
also that one should expect the existence of the
Greisen-Zatsepin-Kuzmin cut-off \cite{Greisen:1966jv,Zatsepin:1966jv}
in the cosmic ray spectrum in this model.

\section{Conclusions}

In this paper we have considered different mechanisms which could
potentially explain the observed correlations of the cosmic ray events
with BL Lacs at the energy $E\sim 10^{19}$ eV and the angle $\sim
0.6^\circ$ coincident with the angular resolution of the HiRes
experiment. We found that the mechanisms which assume only known
particles and interactions under-produce the flux of neutral particles
needed to explain these correlations by at least two orders of magnitude.

There remains a possibility of the astrophysical solution, which is
related to our insufficient knowledge of the Galactic magnetic field.
The observed tight correlations can potentially be explained if there
exist ``windows'' in the Galactic magnetic field with a very low value
of the coherent component of the field and a small coherence length of
the turbulent component. Though this possibility is exotic, it cannot
be ruled out at present.

The mechanisms which we have discussed in this paper are based on the
known physics, i.e. they certainly operate in Nature provided the
cosmic ray flux contains light nuclei or protons. One of these
mechanisms, the conversion of protons to neutrons, 
implies that at energies around $10^{20}$~eV a
few percent of the ultra-high energy protons (cf. Fig.~\ref{fig:F})
get converted into neutrons and cross the Galactic magnetic field
undeflected. Therefore, if the cosmic rays with the energy around the
GZK cutoff are protons, there must be a few percent fraction of them
that point back to the sources with the accuracy better than a
fraction of a degree, provided the extragalactic magnetic fields are
small. With a large statistics, this may allow to measure separately
the Galactic and extra-galactic magnetic fields and to verify by an
independent method the chemical composition of UHECR.

\acknowledgements
{\tolerance=400 We are grateful to A.~Kashlinsky, V.~Kuzmin, D.~Semikoz, 
  G.~Thompson and T.~Weiler for valuable comments and discussions. 
  I.T. thank the Galileo
  Galilei Institute for Theoretical Physics for the hospitality and the INFN
  for partial support during the completion of this work. The work of
  P.T. is supported in part by the IISN, Belgian Science Policy (under
  the contract IAP V/27).}

\end{document}